\documentclass[reprint,superscriptaddress, amsmath,amssymb, aps, pra]{revtex4-2}
\usepackage{soul} 
\usepackage{dcolumn}
\usepackage{bm}
\usepackage{comment}

\usepackage[usenames, dvipsnames]{color}
\usepackage{hyperref}
\hypersetup{
    colorlinks=true,
    linkcolor=Blue,
    citecolor=Blue,
    filecolor=Blue,      
    urlcolor=Blue,
    pdftitle={Vortex Lattices in Strongly Confined Quantum Droplets},
    }
\usepackage{graphicx}

\newcommand{\red}[1]{{#1}}
\renewcommand{\st}[1]{}
\usepackage[caption=false]{subfig}

\begin{document}


\title{Vortex and fractional quantum Hall phases in a rotating  anisotropic Bose gas}
\author{U. Tanyeri}
 \affiliation{%
Department of Physics, Middle East Technical University, Ankara, 06800, T\"{u}rkiye\\
}%

\author{A. Kallushi}
 \affiliation{%
Department of Applied Mathematics and Theoretical Physics, University of Cambridge, Wilberforce Road, Cambridge CB3 0WA, UK\\
}

\author{R. O. Umucalılar}
 \affiliation{%
Department of Physics, Mimar Sinan Fine Arts University, 34380 Sisli, Istanbul, T\"{u}rkiye\\
}%
\author{A. Keleş}
 \affiliation{%
Department of Physics, Middle East Technical University, Ankara, 06800, T\"{u}rkiye\\
}%

\date{\today}

\begin{abstract}
Realizing fractional quantum Hall (FQH) states in ultracold atomic systems remains a major goal despite numerous experimental advances in the last few decades. Recent progress in trap anisotropy control under rapid rotation has renewed interest in ultracold atomic FQH physics, enabling experiments that impart much larger angular momentum per particle and offer in-situ imaging with resolution finer than the cyclotron orbit size. In this paper, we present a theoretical investigation of a rapidly rotating anisotropic Bose gas. By projecting the full Hamiltonian, including both kinetic and interaction terms, onto the lowest Landau level, we derive a compact two-parameter model that captures the effects of interaction strength, rotation rate, and anisotropy. Using exact diagonalization and density matrix renormalization group, we obtain a phase diagram that features broken-symmetry phases and topologically ordered quantum Hall states, while also highlighting the distinctive physics arising from the system’s edges. 
Our results demonstrate the potential for future theoretical and experimental exploration of anisotropic quantum fluids, offering a unified framework 
for weakly interacting Bose condensates, vortex matter, and strongly correlated topological phases.

\end{abstract}

\maketitle



\section{Introduction} \label{sec:intro}

The fractional quantum Hall (FQH) effect is one of the most influential topics in condensed matter physics \cite{prange_girvin_qhe}. The experimental discovery of integer \cite{pepper1980} and fractional \cite{gossard1982} quantization of transverse Hall conductivity and their revolutionary theoretical explanations \cite{tknn, laughlin} led to a paradigm shift in our understanding of phases of matter and phase transitions. 
Topology, geometry, and symmetry beyond the Ginzburg-Landau paradigm have become central in our approach to quantum mechanical systems. Realizing FQH states in ultracold atomic systems remains one of the field’s most sought-after goals \cite{Cooper2008,Viefers2008,Fetter2009} despite numerous groundbreaking experimental achievements—many of which realized condensed matter phenomena once thought impossible \cite{pethick2008}.


Bosonic atomic gases exhibit several distinct characteristics compared to electronic systems in solid-state devices
\cite{Cooper2008,Viefers2008,Fetter2009}. Unlike electrons, atoms in an ultracold gas are neutral and do not couple to magnetic fields. Nevertheless, 
mechanical rotation can be employed instead of magnetic field as the Coriolis force on a massive particle in the rotating frame mathematically mimics the Lorentz force on a charged particle.
Secondly, bosonic gases undergo Bose-Einstein condensation (BEC) at low energies, forming a symmetry-broken state of matter rather than topologically ordered FQH states. 
The macroscopically occupied, single-valued complex wavefunction of the BEC leads to an irrotational, homogeneous superfluid state. This is analogous to the expulsion of magnetic field in the Meissner state of type-I superconductors. When the condensate is rotated above a critical threshold, singular vortices form with quantized circulations. At higher rotation rates, more vortices appear \cite{butts1999} and eventually arrange into a triangular lattice of singularities \cite{ketterle2001,dalibard2004,cornell2004} similar to Abrikosov vortex lattice in type-II superconductors under strong magnetic fields \cite{abrikosov1957}.
%
As the rotation frequency approaches the trap frequency, the system enters a unique state called the mean-field quantum Hall regime, where the condensate’s wavefunction is confined to the lowest Landau level (LLL) manifold, leading to the formation of a dense lattice of vortices \cite{ho2001}.
The vortex state forms an inhomogeneous superfluid with intricate symmetry-breaking properties \cite{murayama2013}, yet it remains distinct from the topologically ordered phases.
%

%
%
At extreme rotation rates, the vortex lattice begins to dissolve as the average interparticle distance approaches the intervortex spacing, making isolated vortices indiscernible
\cite{Aftalion2009,Fetter2009}.
This so-called quantum melting regime of the vortex lattice can also be understood in terms of the Lindemann criterion, which states that melting occurs when the amplitude of individual vortex fluctuations reaches the intervortex spacing \cite{rozhkov1996}. 
The particle and vortex densities can be parametrized by a single dimensionless number $\nu$, the filling factor, defined as the number of particles per vortex. At large filling factors, the system forms a well-defined condensate with a vortex lattice, whereas at low filling factors, it transitions to a strongly correlated quantum liquid.
This picture has been supported through extensive exact diagonalization studies using toroidal geometry, which have demonstrated a quantum phase transition around $\nu\sim 6$  from a vortex lattice ground state, described by the Gross-Pitaevskii formalism, to a FQH phase characterized by a Laughlin-like ground state \cite{Cooper2001}.
Calculations of zero-temperature fluctuations of vortex positions within a Bogoliubov framework have yielded a critical filling factor in agreement with numerical studies \cite{macdonald2002}. However, it remains unclear how these landmark results generalize to condensates in anisotropic traps \cite{Fetter2001, Oktel2004, Fetter2007}, or how finite-size constraints on the vortex lattice \cite{Shlyapnikov2005, Shlyapnikov2009} affect the transition point. Furthermore, the role played by quantum correlations in the presence of anisotropy and edges \cite{Kovrizhin2010,Hu2012} is not yet well understood in the current literature.

Anisotropy and edges offer a fruitful handle to probe many FQH phenomena, previously proposed for ultracold atomic gases in other settings \cite{PhysRevB.69.235309,Cooper1999, Chang2005, Regnault2006}. 
%
In the rapid rotation limit of an anisotropic trap, when the rotation frequency matches the weaker trap frequency, the system maps to a cylindrical geometry in the Landau gauge, with periodic boundaries along the weak axis and open boundaries along the strong axis. Then, the cylinder's circumference, relative to its length, becomes a key parameter \cite{Rezayi1994}.
When the circumference is very large, the condensate forms a narrow channel, and the two edges of the condensate with opposite chiral currents overlap. This so-called pure edge limit provides an ideal setting to study the Luttinger liquid paradigm \cite{Rezayi1994,Bergholtz2009, jolicoeur2012}.
Conversely, when the circumference is small, referred to as the thin cylinder limit, the ground state approaches a density-wave-like state known as the Tao-Thouless state \cite{Tao1983}, where greater analytical control is possible \cite{Bergholtz2005, Seidel2007, Bergholtz2008,hansson2009}.

Recent advances in engineering trap anisotropy have reignited interest in exploring these long-standing FQH problems in ultracold atomic systems via rapid rotation experiments \cite{Fletcher2021, Mukherjee2022, yao2024observationchiraledgetransport}. By smoothly ramping the rotation frequency above the weaker trapping frequency in an asymmetric confinement, the atomic cloud is squeezed into an elongated geometry, effectively occupying a single Landau gauge orbital \cite{Fletcher2021, PhysRevA.105.023310, Andrade2021}. While earlier rapid rotation experiments were limited to imparting angular momentum of up to about 60$\hbar$ per particle \cite{ketterle2001}, recent work with anisotropic traps has achieved values approaching 1000$\hbar$, holding promise for the realization of mesoscopic FQH samples with thousands of atoms \cite{Fletcher2021}. Remarkably, this was achieved without significant atomic loss, even as the interatomic distances expanded to scales comparable to cyclotron orbit sizes. Furthermore, while earlier studies primarily relied on time-of-flight expansion to probe vortices and vortex lattices, the latest experiments enable in-situ imaging of vortex structures with spatial resolution finer than the cyclotron orbit scale.
Subsequent experiments using this setup successfully demonstrated the emergence of chiral edge states, analogous to those in quantum Hall systems \cite{yao2024observationchiraledgetransport}, as well as crystallization phenomena in the strong interaction regime \cite{Mukherjee2022}.
In parallel, significant progress in rotating dipolar gas experiments has revealed phenomena strikingly similar to those observed in asymmetric traps, including elongated density profiles and vortex arrangements---highlighting the essential role played by anisotropy in stabilizing these quantum states \cite{ferlaino2022}.

In this paper, we present a theoretical framework for a rapidly rotating anisotropic Bose gas. Starting from the second-quantized Hamiltonian for a rotating Bose gas with contact interactions in two dimensions, we focus on the rapid rotation limit---where the rotation frequency matches the weak trap frequency---and show how the system maps to the Landau gauge in cylindrical geometry. In this setting, the presence of edges along the open direction of the cylinder plays a crucial role in shaping the system’s properties. By projecting the full Hamiltonian (including both kinetic and interaction terms) onto the LLL, we derive a compact two-parameter model that captures the effects of interaction strength, rotation rate, and anisotropy. This model \st{resembles a tight-binding} \red{effectively describes a one-dimensional quantum} chain with all-to-all long-range interactions and features both conventional conserved quantities, such as the total number of particles in the chain, and unconventional ones, such as the center of mass coordinate—which corresponds to the total momentum along the translationally invariant (periodic) direction. 

Using exact diagonalization (ED) \cite{quspin2017,quspin2019} and density matrix renormalization group (DMRG) methods \cite{itensor,itensor-r0.3}, we uncover a phase diagram that unifies uniform Bose-Einstein condensates, various vortex lattice structures, and strongly correlated phases, while also highlighting the distinctive physics arising from the system’s edges. To identify different phases with broken symmetries and/or topological order, we employ a variety of diagnostic tools, including the particle-hole energy gap, entanglement entropy, and the overlap with the Laughlin wavefunction in cylindrical geometry.

This paper is organized as follows: In Sec.~\ref{sec:model}, we introduce the model Hamiltonian projected to the LLL. In Sec.~\ref{sec:results}, we present details of the phase diagram featuring a uniform Bose gas, various vortex lattices, and FQH states. In Sec.~\ref{sec:conclusion}, we present a discussion of our results and finish with conclusions and outlook.

\section{Model} \label{sec:model}

\st{We consider a single component trapped bosonic gas with a much stronger confinement in the z-direction than in the other two directions.}
\red{We consider a single-component quasi-two-dimensional bosonic gas rotating about the $z$-axis, which is perpendicular to the confinement plane.}
By integrating out the $z$-direction, we obtain a two-dimensional Hamiltonian 
\red{ \cite{Petrov2004-lowD} }
\begin{equation}\label{eqn:H_field}
    \hat H = \int \mathrm d^2 r \red{\Bigg[}\hat \Psi^\dagger(\mathbf{r}) \hat H_0 \hat \Psi(\mathbf{r}) + \frac{g_\text{2D}}{2} \hat{\Psi}^\dagger(\mathbf{r})\hat{\Psi}^\dagger(\mathbf{r})\hat \Psi(\mathbf{r})\hat \Psi(\mathbf{r})\red{\Bigg]},
\end{equation}
where $\mathbf{r}=(x,y)$, $\hat\Psi(\mathbf{r})$ and $\hat\Psi^\dagger(\mathbf{r})$ are bosonic annihilation and creation operators with commutation $[\hat\Psi(\mathbf{r}),\hat\Psi^\dagger(\mathbf{r}')]=\delta(\mathbf{r}-\mathbf{r}')$, $g_\text{2D}=\sqrt{8\pi}\hbar^2a_s/a_zm$ is the effective 2D interaction strength, $a_s$ is the $s$-wave scattering length, $a_z$ is the harmonic oscillator length in the $z$-direction, and $m$ is the mass. The non-interacting Hamiltonian $\hat H_0$ describes a rapidly rotating system with angular velocity $\Omega\hat z$ and anisotropic trapping potential, written in the rotating reference frame as
\begin{equation} \label{eqn:H_0-before_gauge}
    \hat H_0 = 
    \frac{\mathbf{p}^2}{2m}
    +  \frac{1}{2}m\omega_x^2x^2 + \frac{1}{2}m\omega_y^2y^2 -\Omega L_z.
\end{equation}
Here $\mathbf{p}=(p_x,p_y)$ is the 2D momentum, $L_z=xp_y-yp_x$ is the $z$-component of angular momentum, $\omega_x$ and $\omega_y$ are trap frequencies in the $x$- and $y$-directions, respectively. 

\begin{figure}[b]
    \centering
    \includegraphics[width=0.75\linewidth]{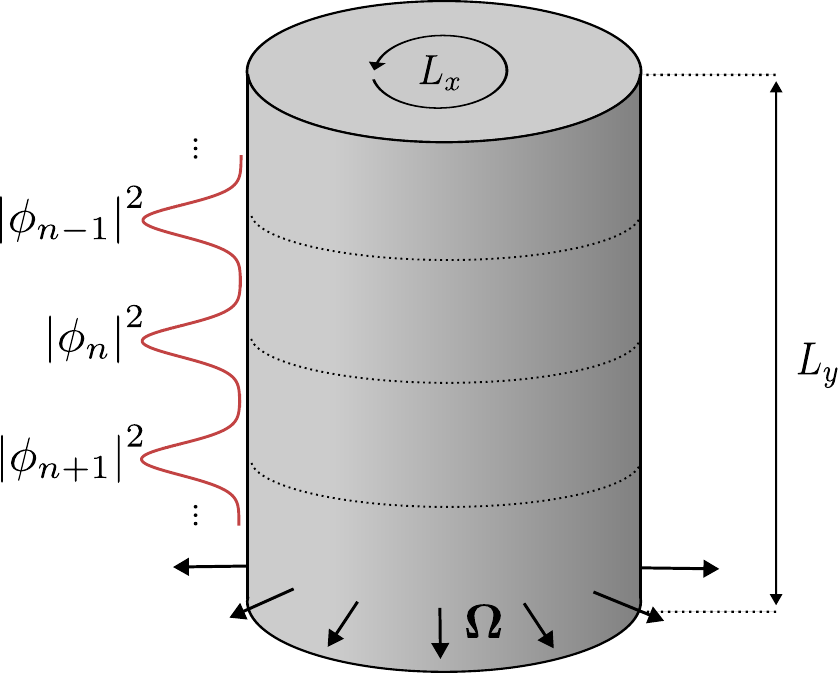}
    \caption{Geometry of the 2D rotating atomic Bose gas in the Landau gauge and schematic representation of the eigenfunctions of the noninteracting Hamiltonian.}
    \label{fig:cylinder}
\end{figure}

In what follows, we focus on the anisotropic rapid rotation limit defined as $\Omega\approx\omega_x$ and $\Omega<\omega_y$ \st{, which gives}\red{. In this limit, we assume the trapping in the $x$-direction vanishes so that our model describes} a 2D system \st{with a trapping} \red{trapped} only in the $y$-direction. Applying a gauge transformation to our field operators $\hat \Psi\rightarrow e^{\mathrm i m\Omega xy/\hbar}\hat \Psi$ leaves the interaction term invariant, but the single-particle Hamiltonian becomes  
\begin{equation} \label{eqn:H_0}
    \hat H_0  = 
    \frac{1}{2m}(p_x + 2m\Omega y)^2 + \red{\frac{p_y^2}{2m}} + \frac{1}{2}m(\omega_y^2-\Omega^2)y^2.
\end{equation}
This is the non-interacting Hamiltonian of the quantum Hall system in the Landau gauge with added trap potential in the $y$-direction, where effective electron charge $q^*$ and magnetic field $B^*$ map to $q^*B^*=2m\Omega$ \cite{Cooper2008}. Using the translational invariance along the $x$-axis thanks to \red{the assumption of} vanishing trapping,  we substitute eigenfunctions of the form $e^{\mathrm i kx} \phi_k(y)$, and assume periodic boundary conditions along the $x$-direction. The allowed wavevectors are $k=2\pi n/L_x$, where $L_x$ is the size of the system in the $x$-direction, as illustrated in Fig.~\ref{fig:cylinder}.
The Hamiltonian $\hat H_k$ for the functions $\phi_k(y)$ can be obtained as \red{\cite{Shlyapnikov2005}}
\begin{equation} \label{eqn:H_k}
    \hat H_k = \frac{p_y^2}{2m} + \frac{1}{2}m\omega_c^2 (y+y_k)^2 + \frac{\hbar^2k^2}{2m_*},
\end{equation} where $\omega_c=\sqrt{3 \Omega^2+\omega_y^2}$, $m/m_*=1-4 \Omega^2/\omega_c^2, \quad y_k=\frac{2 \Omega}{\omega_c} k \ell^2$, and $\ell=\sqrt{\hbar/m \omega_c}$ \cite{Shlyapnikov2005}. This is the quantum harmonic oscillator Hamiltonian centered at $y=-y_k$ with the LLL eigenvalues 
\begin{equation}\label{eqn:eigenvals}
    \varepsilon_k=\frac{\hbar \omega_c}{2}+\frac{\hbar^2 k^2}{2 m^*},
\end{equation} 
and the corresponding normalized eigenfunctions
\begin{equation}\label{eqn:eigenfuns}
\varphi_k(x, y)=\frac{1}{\sqrt{L_x}} e^{i k x} \frac{1}{\left(\pi \ell^2\right)^{1 / 4}} e^{-\frac{1}{2 \ell^2}\left(y+y_k\right)^2}. 
\end{equation}
We take a finite strip of length $L_y$ in the transverse direction as shown in Fig.~\ref{fig:cylinder}, and impose a momentum cutoff by requiring the so-called guiding centers $y_k$ to be inside the sample domain $-L_y/2\le y_k<L_y/2$. Using $k=2\pi n/L_x$ and the definition of $y_k=\frac{2\Omega}{\omega_c}k\ell^2$ written after Eq.~\ref{eqn:H_k}, this leads to 
\begin{equation}
    -\frac{N_v}{2}\le n <\frac{N_v}{2},
    \label{eqn:n_range}
\end{equation}
where \st{the ``number of vortices'' in the system is defined as}
\begin{equation}
    N_v= \frac{L_xL_y}{2\pi\ell^2}\frac{\omega_c}{2\Omega}.
    \label{eqn:number_of_vortices}
\end{equation}
\st{This quantity}\red{In the isotropic limit $\omega_c\rightarrow2\Omega$, this quantity becomes the standard definition of the number of vortices in a superfluid\cite{Cooper2008} and} takes place of the number of magnetic flux quanta in the electronic quantum Hall problem. \red{Thus, from now on, we refer to $N_v$ as the ``number of vortices'' in the system.}

We assume that the Landau level spacing $\hbar\omega_c$ is much bigger than other energy scales in the system, and project the Hamiltonian to the LLL using the expansion of the field operators
\begin{equation}
 \hat{\Psi}(x,y) = \sum_n \hat a_n \varphi_n(x,y)   
\end{equation}
where $\hat a_n$ and $\hat a_n^\dagger$ are canonical bosonic annihilation and creation operators in state $\varphi_n$ with commutation $[\hat a_n,\hat a_{n'}^\dagger]=\delta_{n,n'}$, and we re-label the functions in \eqref{eqn:eigenfuns} with integers $n$ using the relation $k=2\pi n/L_x$. Hence, the full many-body Hamiltonian takes the following form in the LLL manifold \red{\cite{Shlyapnikov2005}}
\begin{equation} \label{eqn:Hamiltonian}
    \hat H = \sum_n n^2 \hat a^\dagger_n \hat a_n  
    + 
    \sum_{n_1,n_2,n_3}  V_{n_1,n_2,n_3} \hat a^{\dagger}_{n_1} \hat a^{\dagger}_{n_2} \hat a_{n_3} \hat a_{n_4},
\end{equation} 
where $n_4=n_1+n_2-n_3$,
and the interaction matrix elements are written as
\begin{equation}
    V_{n_1,n_2,n_3} = g e^{-2 \pi^2 \gamma^2\left[(n_1-n_3)^2+(n_2-n_3)^2\right]},
    \label{eqn:interaction}
\end{equation}  
with the dimensionless interaction strength given by
\begin{equation} 
    g=\frac{1}{2\pi^2}\frac{L_x}{\ell} \frac{m_*}{m}\frac{a_s}{a_z}.
    \label{eqn:g}
\end{equation} 
In Eq.~\eqref{eqn:Hamiltonian}, we rescale the total Hamiltonian with energy $\frac{\hbar^2}{2m_*}\frac{4\pi^2}{L_x^2}$, drop the overall constant $\frac{1}{4\pi^2}\frac{m_*}{m}\frac{L_x^2}{\ell^2}$ from the quadratic term, which corresponds to the zero-point energy of the LLL, and define 
\begin{equation} 
\gamma = \frac{\ell}{L_x} \frac{2\Omega}{\omega_c}.
\label{eqn:gamma}
\end{equation}
This expression is renormalized by the trapping factor $2\Omega/\omega_c$ with respect to the earlier definition in Ref.~\cite{Rezayi1994}. Note that one can also define so-called aspect ratio $a=2\pi\gamma^2 N_v=\frac{L_y}{L_x}\frac{2\Omega}{\omega_c}$, which has the same factor $2\Omega/\omega_c$ compared with the one in Ref.~\cite{Rezayi1994}, but we will use $\gamma$ as a parameter to scan the phase diagram. The Hamiltonian in Eq.~\eqref{eqn:Hamiltonian} conserves the total number of particles, which can be written as
\begin{equation}
    N = \sum_n \hat a^\dagger_n\hat a_n.
    \label{eqn:number_of_particles}
\end{equation}
Furthermore, due to the translational invariance along the $x$-direction, total momentum along $x$ is also conserved, which we write in dimensionless form as
\begin{equation}
    K = \sum_n n \hat a^\dagger_n\hat a_n.
    \label{eqn:x_momentum}
\end{equation}

The rapidly rotating anisotropic 2D many-body problem given in Eq.~\eqref{eqn:H_field} is reduced to a 1D ``\st{tight-binding}\red{quantum} chain'' \red{as} in Eq.~\eqref{eqn:Hamiltonian} with ``onsite potential'' $n^2$ given \red{by} the first term and all-to-all coupling ``long-range interaction'' given \red{by} the second \red{one}. 
The parameters of the first model are the trap frequencies $\omega_x$, $\omega_y$, the rotation speed $\Omega$, where we only consider the limit $\Omega\rightarrow\omega_x$, and the interaction strength $g_\text{2D}$. This rich set of parameters is mapped to two parameters in \st{the tight-binding} \red{our} model: the anisotropy $\gamma$ and interaction $g$. To fully specify the many-body problem for a given number of particles $N$, one also has to determine the vortex number $N_v$, which acts like the total number of sites in the \st{tight-binding} chain and depends on $L_x$ and $L_y$.
We will not derive a direct relationship between these lengths and trap frequencies (e.g. as in Refs.~\cite{Fetter2009,Aftalion2009}), which is outside the scope of this paper, and look instead for quantum phases as a function of parameters $g$ and $\gamma$ of the reduced model. 
Also note that the Hamiltonian projected to the LLL subspace does not depend on $\Omega$ and $g_\text{2D}$ individually, which necessitates the definition of $g$ in Eq.~\eqref{eqn:g} depending on the anisotropy, interaction strength, and the rotation rate. Previously, Ref.~\cite{Aftalion2005} also reported a similar observation via the minimization of the Gross-Pitaevskii action in the LLL. Nevertheless, we will refer to $g$ as the interaction constant below for brevity.  
In addition, given the total number of bosons $N$ occupying the sites $N_v$, the key quantity in the FQH problem is the so-called filling factor 
\begin{equation}
    \red{\nu = \frac{N}{N_v},}
\end{equation}
which corresponds to the number of particles per vortex.
Interestingly, whereas the conservation in Eq.~\eqref{eqn:number_of_particles} corresponds to the usual total number of particles occupying the \st{tight-binding} \red{sites in the} chain, the momentum conservation in Eq.~\eqref{eqn:x_momentum} corresponds to the conservation of the ``dipole moment'' or the center of mass of the chain. In passing, we note that such models with unconventional conservations received renewed interest in recent literature (see Ref.~\cite{knap2024,principi2024,pollmann2024} and references therein for recent examples and Ref.~\cite{simon1998,moore2005} for the FQH context). 

It is instructive to consider the isotropic limit $\omega_x=\omega_y=\Omega$ in our problem \footnote{this limit should be taken before rescaling with the full Hamiltonian with $\frac{\hbar^2}{2m_*}\frac{4\pi^2}{L_x^2}$}, which gives $\omega_c=2\Omega$ and $m/m_*=0$. This cancels the first term in Eq.~\eqref{eqn:Hamiltonian} and leaves only the interaction, as expected. However, our system is still inherently anisotropic due to distinct boundary conditions in the $x$- and $y$-directions, which makes it impractical to study a rotationally symmetric system. This observation should be contrasted with the analytical subtlety of including anisotropy in the symmetric gauge within the LLL description of a rapidly rotating system \cite{Oktel2004, Fetter2007}. One can consider the toroidal geometry to eliminate boundaries as in \cite{Cooper2001}, but this will not be pursued here since we are interested in a finite anisotropic system with nontrivial edge effects that could be more relevant for the recent experiments \cite{Fletcher2021, Mukherjee2022,yao2024observationchiraledgetransport}.

The model given in Eq.~\eqref{eqn:Hamiltonian} describes a rich variety of quantum phases and their intriguing competition. In the limit of small $g$, which corresponds to weak two-body interactions or slow rotation, the Bogoliubov approximation can be employed to demonstrate Bose-Einstein condensation and its excitations. The resulting spectrum exhibits the well-known superfluid dispersion characterized by the so-called roton-maxon structure \cite{Shlyapnikov2005}. \red{This phase is the conventional uniform Bose-Einstein condensate without any vortex.} Below, this phase is referred to as Meissner state, based on the terminology established in quasi-one-dimensional ultracold atomic gases in recent years \cite{bloch2014}. As $g$ increases, the roton minimum softens, and beyond a critical value, the ground state undergoes a phase transition from a uniform superfluid to a vortex state.
In the isotropic rapid rotation limit $\Omega\rightarrow\omega_x=\omega_y$, which corresponds to large $g$ in our model, the condensate acquires a large number of vortices with triangular lattice structure \cite{Fetter2009}. 
Beyond another critical value where cores of vortices start to overlap, and the number of vortices becomes of the order of the number of particles \cite{Cooper2001}, the strong quantum fluctuations melt the vortex lattice, giving way to correlated FQH states described by the Laughlin wavefunction \cite{Rezayi1994, Bergholtz2008}. 
Below, we assess the validity of this scenario and investigate the competition between these phases quantitatively using ED of a small system ($N=5$) \cite{quspin2017,quspin2019} and 
use these results to benchmark our findings for larger systems ($N=5$, $8$, $10$) obtained using DMRG \cite{itensor}.

\red{While the effective model we study is based on earlier work like Ref.~\cite{Shlyapnikov2005}, our treatment retains full dependence on anisotropy through the effective mass and trap asymmetry, and reduces the problem to two key parameters, 
$g$ and $\gamma$ even for the case $\omega_y\neq\Omega$. This allows us to systematically explore the entire range from weak to strong interactions and from isotropic to highly anisotropic geometries within a unified framework. Unlike prior work on isotropic uniform limit \cite{Cooper2001}, we access edge-dominated regimes, identify crossover behavior unique to strong anisotropy, and uncover unique transitions—including a Meissner-to-FQH crossover in the pure edge limit—not captured in earlier studies.}

\section{Results} \label{sec:results}

\begin{figure}[t]
    \centering
    \includegraphics[width=0.95\linewidth]{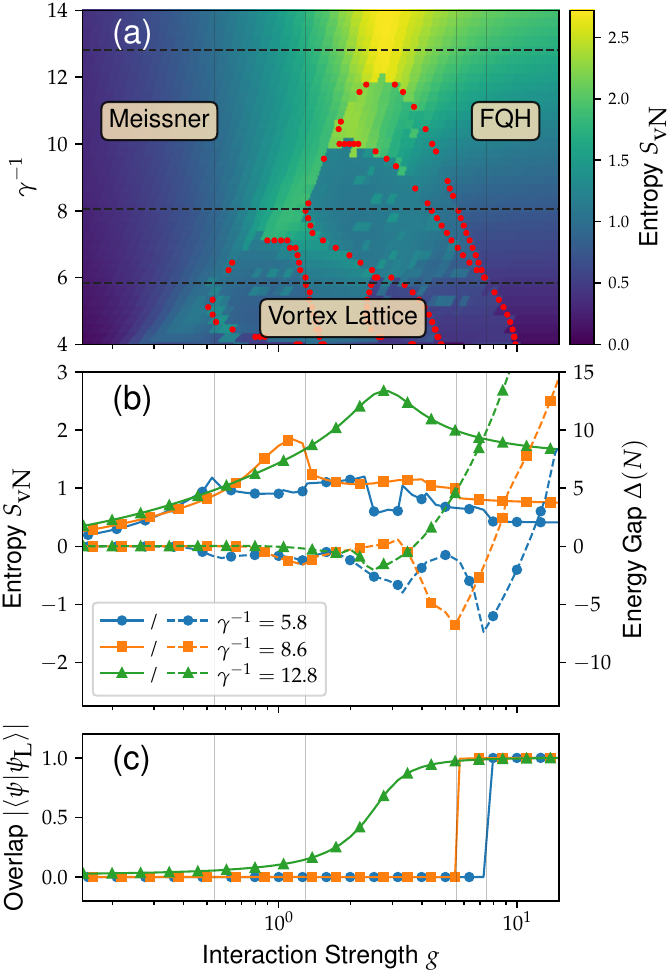}
    \caption{(a) The phase diagram of a rapidly rotating anisotropic Bose gas  described by the Hamiltonian \eqref{eqn:Hamiltonian} for $N=5$ as a function of the interaction strength $g$ 
    and anisotropy $\gamma$.
    The color bar indicates the von Neumann entanglement entropy calculated with DMRG.
    The red dots are obtained from ED
    by finding the change of the ground state momentum (see the main text). 
    \red{For $\gamma^{-1}<4$, inter-site interactions are exponentially suppressed.}
    (b) The von Neumann entropy (solid line) and particle-hole excitation gap \eqref{eqn:many_body_gap} (dashed line) along the horizontal cuts in 
    the phase diagram.
    (c) The overlap between the ground state calculated with ED and the $\nu=1/2$ Laughlin state along the same cuts.
    }
    \label{fig:phase_diagram_N_5}
\end{figure}

We focus on the half-filling $\nu=1/2$ by taking $N=5$ bosons occupying $N_v=10$ vortices. We set the maximum occupancy of each site to be the total number of bosons $N$, which is crucial for observing superfluid phases in the numerics. \red{We characterize the ground state using three complementary diagnostics: (i) the von Neumann entanglement entropy, (ii) the particle-hole excitation gap, and (iii) the overlap with the $\nu=1/2$ Laughlin wavefunction. Our results are presented collectively in Fig.~\ref{fig:phase_diagram_N_5}, which serves as a central summary of our findings. Fig.~\ref{fig:phase_diagram_N_5}(a) shows the phase diagram with the entanglement entropy displayed in the color scale. Red dots mark transitions identified via the ED ground-state momentum. Fig.~\ref{fig:phase_diagram_N_5}(b) shows the cuts of entanglement entropy and particle-hole gap across selected anisotropy values, indicating phase boundaries. Fig.~\ref{fig:phase_diagram_N_5}(c) shows the overlap with the Laughlin wavefunction, signaling the emergence of FQH order. These observations jointly distinguish the Meissner, vortex lattice, and FQH phases. We now discuss each phase and our methodology in detail.}

The full phase diagram is presented in Fig.~\ref{fig:phase_diagram_N_5}(a). 
The phase boundaries are initially determined using DMRG via the calculation of the von Neumann entanglement entropy $S_\text{vN}$ by cutting the system in two halves with almost equal number of sites in each subsystem.
$S_\text{vN}$ serves as a useful indicator of a second-order phase transition, as the correlation length diverges at the critical points. This leads to sharp changes in the entanglement entropy even for small systems, as demonstrated in Fig.~\ref{fig:phase_diagram_N_5}(b). 
\st{To confirm these boundaries independently, we compute the expectation value of the total momentum [Eq.~(15)] in the ground state with ED and mark the boundaries with red dots in Fig.2(a). As an example, along the cut $\gamma^{-1}\approx6$ in the figure, the ground state momentum follows the sequence: $K=0$, then $K=1$, back to $K=0$, a small region with $K=2$, followed by $K=1$ and finally $K=0$ again.}
\red{To identify the Meissner–vortex phase boundary, we compute the expectation value of the total momentum [Eq.~\eqref{eqn:x_momentum}] in the ground state using the ED. The boundary is defined by the point at which the 
$K=0$ ground state becomes energetically unfavorable compared to a state with 
$K\neq 0$; these points are marked with red dots in Fig.~\ref{fig:phase_diagram_N_5}(a). We have verified the robustness of the phase boundary via the finite-size studies of entanglement entropy.}
Positive and negative nonvanishing momenta exhibit degenerate energies, manifesting the translational symmetry breaking induced by the vortex lattice ground state. A similar phenomenon is inferred from the emergence of degenerate energy levels at Haldane momenta in torus geometry \cite{Cooper2001}. By calculating the expectation value of density in the ground state, we have obtained sound evidence that each domain exhibits a different vortex lattice. We expect that kinks or abrupt changes in these results reliably indicate the transition between Meissner and vortex phases. However, finite-size effects may prevent an accurate characterization of the competition between distinct lattice structures. This also explains the noisy behavior of the entanglement entropy in the vortex phases seen in Fig.~\ref{fig:phase_diagram_N_5}(b), which is possibly due to the frustration of the lattice by the finite-size constraint. Therefore, we do not pursue an investigation of the competition between distinct vortex lattice structures in the thermodynamic limit, which can be addressed reliably with a Gross-Pitaevskii formalism \cite{keles2025}. \red{In the thermodynamic limit, there also exist powerful analytical methods which can accurately determine various vortex lattice geometries \cite{mueller2002}.}

Above $\gamma^{-1}\approx 12$, the phase boundary between Meissner and FQH phases shows a smooth transition in entanglement entropy without any kink, based on the largest available data with $N=5$, $8$, and $10$. This suggests the possibility of a crossover rather than a true phase transition. However, in this regime, the system becomes strongly one-dimensional with arbitrarily long-range coupling (because of large $\gamma$), leading to significantly slower convergence in the DMRG algorithm, which hinders our ability to obtain more conclusive evidence. Below, we discuss this transition in more detail from the perspective of the FQH phase and the Laughlin ansatz.

%
%
%
We calculate the energy gap for particle-hole excitations to probe the compressible superfluid states, which can readily be obtained using both ED and DMRG. In the Meissner and vortex-lattice phase, these excitations are expected to be gapless in the thermodynamic limit due to the presence of a condensate. In contrast, correlated incompressible states, including FQH states and Mott insulators, exhibit a finite particle-hole excitation gap. This gap corresponds to the jump in the chemical potential when adding or removing a particle. In order to minimize the finite-size effects, the energy gap is defined as \red{\cite{Cooper2001}}
\begin{equation} \label{eqn:many_body_gap}
    \Delta(N) = N\left(\frac{E(N+1)}{N+1}+\frac{E(N-1)}{N-1}-2\frac{E(N)}{N}\right),
\end{equation} 
where $E(N)$ is the ground state energy of the Hamiltonian \eqref{eqn:Hamiltonian} for a system with $N$ bosons. 
As shown in Fig.~\ref{fig:phase_diagram_N_5}(b), the energy gap $\Delta(N)$ remains small in both Meissner and vortex lattice phases, albeit nonvanishing due to combined effects of the trap potential, the finite particle number, and the edges. However, upon entering the FQH phase, the energy gap increases proportionally to the interaction strength $g$, signaling the onset of the incompressible ground state, characteristic of the FQH phase \cite{Cooper2001}. 

\begin{figure}
    \centering
    \includegraphics[width=0.45\textwidth]{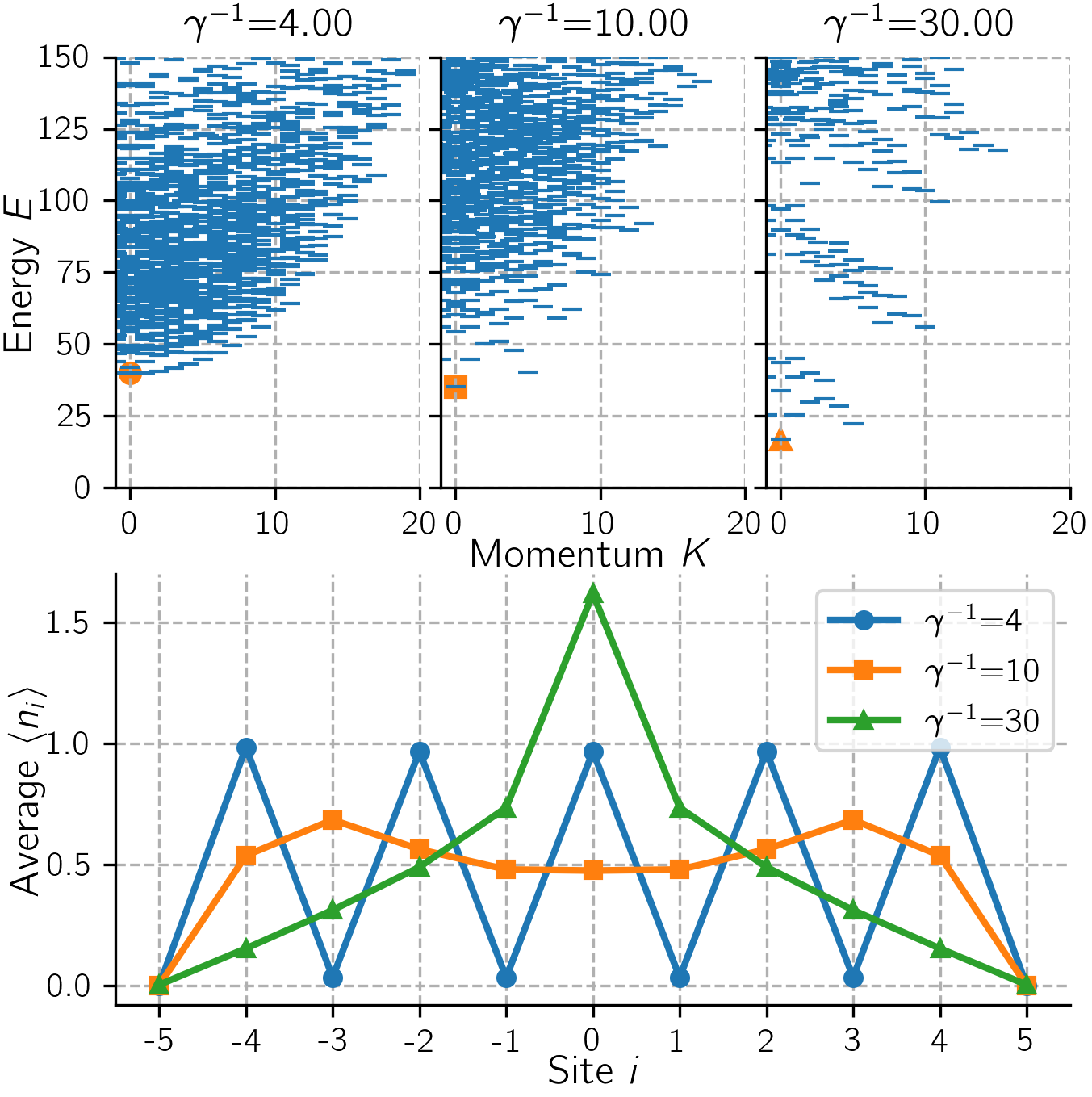}
    
    %
    \caption{(Top) Energy spectrum of the Hamiltonian \eqref{eqn:Hamiltonian} for $N=5$ bosons with $g=10$ and $\gamma^{-1}=4$, $10$ and $30$, which is cut through the FQH phase in Fig.~\ref{fig:phase_diagram_N_5} vertically. We choose the number of sites to be 11 to have a symmetric spectrum (density) with respect to the origin $K=0$ ($i=0$). Ground state is labeled with (orange) marks for better visibility. (Bottom) Occupations $\langle n_i\rangle$ in the ground state for the trap minima located in the middle site which shows transition from the Tao-Thouless state to the Laughlin state and then to the quasi-1D state, as the anisotropy ratio $\gamma$ tuned from thin-cylinder to two-dimensional and then to wide-cylinder limit regimes.}
    \label{fig:cut_through_fqh}
\end{figure}

In the strong interaction limit, $g\gg1$, the influence of the trap potential becomes negligible, and a strongly correlated ground state 
emerges, whose nature is determined solely by the anisotropy parameter $\gamma$ \cite{Tao1983, Rezayi1994,Bergholtz2008}. 
When $\gamma^{-1}\ll1$ \footnote{\red{More precisely, the rapid decay of the interaction parameters with $\gamma$ in this limit can be established for $2\pi^2\gamma^2>1$, which implies $\gamma^{-1}<4.4$ but we refer to this limit as $\gamma^{-1}\ll 1$ for simplicity}}, the cylinder's circumference $L_x$ becomes much smaller than its length $L_y$, a regime known as the thin-cylinder limit.
In this limit, the interaction matrix elements in Eq.~\eqref{eqn:interaction} exhibit a
sharp exponential decay, leading to a strong but short-range repulsion, which is minimized by placing the atoms on every other site. As a result, the ground state forms a ``crystal" of atoms with equal spacing, known as the Tao-Thouless state 
\cite{Tao1983,Bergholtz2008}. For instance, at half-filling ($\nu=1/2$), the atomic occupations follow a regular pattern $101010\hdots$. 
In Fig.~\ref{fig:cut_through_fqh}, we present the ED results for a system of 5 bosons on 11 sites,
which manifest such a Tao-Thouless state
for $\gamma^{-1}\approx 4$ and $g=10$.
We have also verified that the low-lying quasiparticle excitations emerge through the creation of a domain wall structure $\cdots0101|1010\cdots$ 
which has been established in the standard Tao-Thouless theory.
%

When $\gamma^{-1}$ increases such that $\gamma^{-1} \lesssim N_v$, the cylinder's circumference $L_x$ and its length $L_y$ become comparable and the system enters a fully two-dimensional regime. Here, longer-range components of the interaction [Eq.~\eqref{eqn:interaction}] become significant and give rise to the melting of the crystalline Tao-Thouless state. This favors the formation of an incompressible Laughlin liquid characterized by a uniform density and the absence of broken translation symmetry.
In Fig.~\ref{fig:cut_through_fqh}, we obtain such a ground state through ED for $\gamma^{-1}=10$ and $g=10$, which reveals uniform $\nu=1/2$ liquid in the center,  which is modified towards the edges as expected.
%

Lastly, when $\gamma^{-1}> N_v$, the cylinder's circumference becomes much larger than its width, resembling a thin ribbon or hoop. In this regime---known as wide-cylinder limit---the system effectively reduces to one-dimension. As $\gamma\rightarrow 0$, the exponential in the interaction [Eq.~\eqref{eqn:interaction}] approaches one, allowing each site to couple to every other site in the chain regardless of distance. Simultaneously, the spacing between the neighboring Gaussian centers, $y_k\propto\gamma$, illustrated in Fig.~\ref{fig:cylinder}, diminishes, and the two opposite chiral edges start to overlap, which brings the system to a pure edge limit without bulk. 
Here, because of the long-range interactions, the correlations in the ground state increase beyond area-law entanglement, and DMRG requires a much bigger bond dimension for convergence.
Nonetheless, this limit can be described by using the hydrodynamic approach and Luttinger liquid formalism \cite{haldane1992,Rezayi1994,wen2004}.  

\begin{figure}
    \centering
    \includegraphics[width=0.9\linewidth]{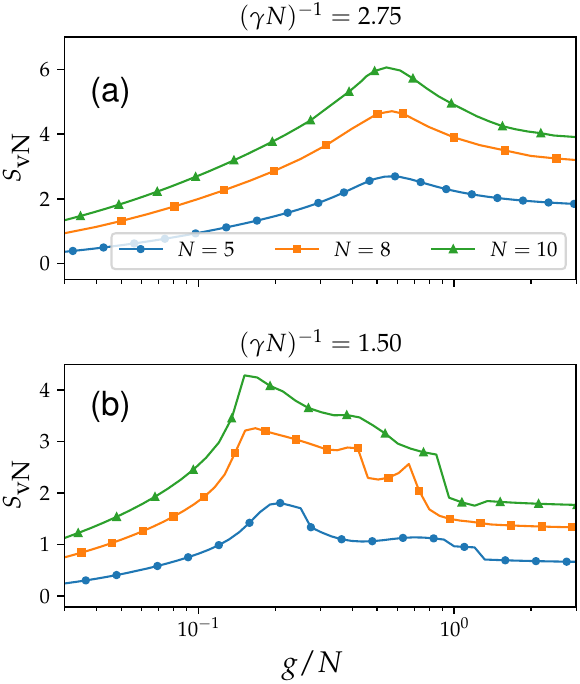}
    \caption{\red{Entanglement entropy $S_\text{vN}$ as a function of the interaction strength $g$ for $N=5$, $8$, and $10$ bosons in the half filling case $\nu=1/2$ using  (a) $\gamma^{-1}=1.375N_v=2.75N$ and (b) $\gamma^{-1}=0.75N_v=1.5N$. In the pure edge limit (a), the scaling with $g/N$ yields a critical point $g_c/N\approx 0.5$ for crossover between the Meissner and FQH phases, and the entropy exhibits a smooth behavior, scaling with $N$. In the smaller $(\gamma N)^{-1}$ limit, the system hosts a vortex lattice phase between the Meissner and FQH phases. The continuous phase transition between the Meissner and vortex lattice phases manifests itself in the cusp of $S_\text{vN}$, which becomes more pronounced as the system size increases.}}
    \label{fig:entropy_vs_g}
\end{figure}

In the \red{isotropic} limit $2\Omega/\omega_c=1$, the first term in Eq.~\eqref{eqn:Hamiltonian} can be dropped, and the resulting Hamiltonian becomes purely interacting. In this case, the Laughlin wavefunction becomes the exact ground state for the present contact interactions, which has been extended to the cylindrical geometry by Thouless \cite{thouless1984theory} (see also \cite{jansen2009symmetry,westerberg1993quantum} for detailed discussions):
\begin{equation} \label{eqn:laughlin-cylinder}
   \psi_\text{L}(x,y)= \prod_{i>j}\left(e^{\mathrm{i}2\pi\gamma z_i/\ell}-e^{\mathrm{i}2\pi \gamma z_j/\ell}\right)^m e^{-\sum_i\frac{y_i^2}{2\ell^2}}.
\end{equation}
where $z=x+iy$.
%
%
We compute the overlap between the ground state obtained from ED and the corresponding Laughlin state with $\nu=1/2$ or $m=2$ along constant-$\gamma^{-1}$ cuts in the phase diagram.
As shown in Fig.~\ref{fig:phase_diagram_N_5}(c), the overlaps jump from zero to nearly unity at a critical interaction strength $g_c$ in both the thin-cylinder and two-dimensional regimes. This transition closely aligns with the vortex lattice-FQH phase boundary previously identified through the energy gap and the entanglement entropy. 
In contrast, in the pure edge limit ($\gamma^{-1}>10$), the overlap evolves smoothly from the Meissner phase to the FQH phase without any sharp kink.  Moreover, the system does not go through an intermediate vortex phase, which is sensible because a one-dimensional fluid cannot accommodate vortices. 
%
At the crossover between these two phases---%
where the overlap with the Laughlin wavefunction increases continuously---the particle-hole gap $\Delta(N)$ does not increase abruptly, signaling 
a smooth evolution of the ground state. However, the von Neumann entropy exhibits a cusp-like structure as shown in Fig.~\ref{fig:entropy_vs_g} for $N=5$, $8$, and $10$ bosons. While the Meissner phase is adiabatically connected to a matrix product state with $S_\text{vN}=0$, the FQH phase shows nontrivial entanglement; thus, a critical point separating these two limits is reasonable.
Note that Eq.~\eqref{eqn:Hamiltonian} in the limit $\gamma^{-1}\rightarrow \infty$  maps exactly to a one-dimensional Bose gas on a ring with contact interactions---the well-known Lieb-Liniger model \cite{lieb1963}. In this case, Ref.~\cite{ueda2003}  reported a crossover from a uniform condensate to a fragmented one (or a soliton state) as a function of the interaction strength based on ED, which is supported by our result shown in Fig.~\ref{fig:entropy_vs_g}. 
A similar phenomenon was considered in Ref.~\cite{huihu2001} using the symmetric gauge, 
which reported a transition to the usual condensed state 
under asymmetric perturbations that break the conservation of total angular momentum. 
To our knowledge, this phenomenon has not been previously studied in the context of FQH phases of rapidly rotating anisotropic gases. 
We anticipate a similar competition with the condensed state in our case when some terms are added to the Hamiltonian that do not conserve the total momentum $K$, such as the trapping along the circumference of the cylinder, which is the subject of future investigation.

\red{The evolution of the ground and excited states in the spectrum in Fig.~\ref{fig:cut_through_fqh} highlights the intricate interplay between interaction and trap-induced anisotropy. In the TT regime, the crystal-like ground state exhibits domain wall and center-of-mass excitations shaped by the underlying quadratic trap and implicit confinement due to the cutoff in $N_v$ \cite{carusotto2017}. As anisotropy decreases, the system transitions into a Laughlin-like FQH liquid, where the excitations include quasiparticles and collective center-of-mass shifts. In the thin-hoop (wide-cylinder) limit, the spectrum becomes highly fragmented, as shifting the central density incurs a large energy cost due to the relocation of a macroscopically occupied state.}

\section{Conclusion} \label{sec:conclusion}

In this paper, we developed a theoretical framework for rapidly rotating anisotropic Bose gases with asymmetric trapping $\omega_x<\omega_y$. The rotation frequency is set to the weaker trapping frequency, $\Omega=\omega_x$, whereas in the orthogonal direction, a residual confinement remains due to $\Omega<\omega_y$. In this setting, the elongated cloud is described by the Landau gauge on a cylindrical geometry where translationally invariant circumference corresponds to trap-free $x$-direction. This geometry is widely studied in the two-dimensional electronic systems under a strong magnetic field, but its bosonic counterpart remains relatively underexplored. In particular, the interplay between Bose condensation, vortex lattice formation, and the strongly correlated phases in this geometry is still an open and intriguing problem.

By projecting the Hamiltonian onto the lowest Landau level in the fast rotation limit, we derived an effective model characterized by two parameters: the interaction strength $g$ and the anisotropy ratio $\gamma$. Using ED, we mapped out the full phase diagram at half-filling ($\nu=1/2$) and identified three distinct regimes: (i) weakly interacting Bose gas without vortices, which we refer to as the Meissner phase, (ii) a series of vortex phases exhibiting different lattice structures depending on $g$ and $\gamma$, and (iii) a strongly correlated phase well described by the Laughlin wavefunction. We benchmarked our main findings in larger systems using the DMRG.

Due to the competition of superfluid phases with broken $U(1)$ and translation symmetries and strongly correlated incompressible phases with topological order, we demonstrated the necessity of employing a variety of complementary probes across the full phase diagram. The compressible superfluid phases were identified through the particle-hole energy gap, while the topologically ordered FQH phase was revealed via the overlap between the exact diagonalization ground state and the Laughlin wavefunction. Interestingly, the entanglement entropy was found to provide a robust diagnostic throughout the entire phase diagram, offering reliable identification of the phase transitions.

In the strongly correlated regime $g\gg1$, the entire range---from wide-cylinder to two-dimensional to thin-cylinder regime---is described by the Laughlin wavefunction. Indeed, Rezayi and Haldane used this fact to construct the phase diagram starting from $\gamma=0$ and building Laughlin wavefunctions for different anisotropy ratios \cite{Rezayi1994}. In our work, neither DMRG nor ED reveals a sharp phase boundary between the Laughlin phase, the Tao-Thouless limit, or the wide-cylinder regime; rather, these limits appear to be adiabatically connected in broad agreement with Ref.~\cite{Bergholtz2008}. While the thin-cylinder limit is analytically tractable and has recently benefited from the development of powerful DMRG-based numerical methods \cite{zeletel2013,zeletel2015}, the wide-cylinder regime remains comparatively less explored and also technically more challenging. To what extent the FQH phase survives in this limit is still an open question \cite{Rezayi1994}, possibly competing with a correlated vortex liquid without condensation \cite{Fetter2009}. Interestingly, recent experiments on cold atomic gases \cite{Fletcher2021, Mukherjee2022, yao2024observationchiraledgetransport} are capable of probing precisely this regime, potentially paving the way for novel insights in the near future.

\red{In experiments \cite{Fletcher2021,Mukherjee2022,yao2024observationchiraledgetransport}, the reported magnetic length is approximately $1.6\mu m$. {\it In situ} images reveal the cloud sizes of $L_x=10-20\mu m$ and $L_y=20-40 \mu m$.
Using these values for the fast rotation regime $\omega_c/2\Omega\approx 1$ yields $N_v=12-50$, which is reasonably close to  the $N_v=10$ value used in our paper. The lowest filling factors reported in \cite{yao2024observationchiraledgetransport}, however, was around $\sim$6, which is far from the necessary half-filling.
A major challenge for next-generation experiments will be achieving such ultradilute regimes while maintaining high-resolution imaging capabilities necessary for reliable observation.
Note that, necessary low-filling regime has recently been achieved in a very different setup, namely a few particles in a small optical lattice experiencing artificial magnetic field created not by rotation but through laser-assisted tunneling \cite{leonard2023realization}.  In that context,  beyond the challenge of reaching the many-body regime with more than a few particles, heating is a concern due to combined effects of optical lattice, Raman beams and Floquet driving.}

Many open questions remain that warrant further investigation. First, the extent to which the topology of the phase diagram persists in the regime $\omega_x<\Omega<\omega_y$---
relevant for current experiments \cite{Mukherjee2022}---is an important issue that will be addressed in a separate study. In this regime, determining a (meta)stable ground state is a delicate theoretical problem due to the inherent instability of the saddle-shaped trap potential. Furthermore, the role of higher Landau levels may be important for explaining some of the observations in these experiments, and a detailed theoretical analysis is required. In the present stable anisotropic case, understanding the exact vortex lattice structures in the thermodynamic limit and its excitations may shed light on the microscopic mechanism behind the vortex lattice melting. Last but not least, rapidly rotating dipolar Bose gases and their low-energy physics in the Landau gauge may offer a compelling comparison to the phase diagram presented here, where the source of anisotropy originates not from the trap potential as in our work but from the two-body interaction itself \cite{ferlaino2022}. 
In summary, our results establish a comprehensive framework that unifies weakly interacting Bose condensates, vortex matter, and strongly correlated topological phases, laying the groundwork for future theoretical and experimental investigations of anisotropic quantum fluids.

\acknowledgments
We thank M. \"O. Oktel for helpful discussions. This work is supported by the Scientific and Technological Research Council of Turkiye (TUBITAK)
1001 program Project No. 124F122 (AK). U.T. is partially supported by the M.Sc. scholarship TUBITAK 2210.

\bibliography{main}

\end{document}